\documentclass[preprint,trackchanges]{aastex61}

\shorttitle{Gammay-ray Variability of LAT AGNs}
\shortauthors{Kushwaha etal. 2017}

\usepackage{multirow}
\begin{document}

\title{Gamma-ray Flux Distribution and Non-linear behavior of Four LAT Bright AGNs}

\author{Pankaj Kushwaha}
\affiliation{Department of Astronomy (IAG-USP), University of Sao
Paulo, Sao Paulo 05508-090, Brazil}
\affiliation{Inter-University Center for Astronomy and Astrophysics, Pune 411007,
India}

\author{Atreyee Sinha}
\affiliation{Inter-University Center for Astronomy and Astrophysics, Pune 411007,
India}

\author{Ranjeev Misra}
\affiliation{Inter-University Center for Astronomy and Astrophysics, Pune 411007,
India}

\author{K. P. Singh}
\affiliation{Department of Astronomy \& Astrophysics, Tata Institute of Fundamental
Research, Mumbai 400005, India}

\author{E. M. de Gouveia Dal Pino}
\affiliation{Department of Astronomy (IAG-USP), University of Sao
Paulo, Sao Paulo 05508-090, Brazil}

\correspondingauthor{Pankaj Kushwaha}
\email{pankaj.kushwaha@iag.usp.br}

\begin{abstract}
We present a statistical characterization of  the $\gamma$-ray emission from the four
\emph{Fermi}-LAT sources: FR I radio galaxy NGC 1275, BL Lac Mrk 421, FSRQs B2
1520+31 and PKS 1510-089 detected almost continuously over a time integration of
3-days between August 2008 - October 2015. The observed flux variation is large,
spanning $\gtrsim 2$ orders of magnitude between the extremes except for Mrk~421.
We compute the flux distributions and compare with Gaussian and  lognormal ones.
We find that the 3 blazars have distribution consistent with a lognormal, suggesting
that the variability is of a non-linear, multiplicative nature. This is further
supported by the computation of the flux-rms relation, which is observed to be
linear for the 3 blazars. However, for NGC 1275, the distribution does not seem to
be represented either by a  lognormal or a Gaussian, while its flux-rms relation
is still found to be linear. We also compute the power  spectra, which suggest
the presence of a break, but are consistent with typical scale-free power-law shot
noise. The results are broadly consistent with the statistical properties of the
magnetic reconnection powered minijets-in-a-jet model. We discuss other possible
scenarios and implications of these observations on jet processes and
connections with the central engine.

\end{abstract}


\section{INTRODUCTION} \label{sec: intro}
Active galactic nuclei (AGNs) refer to astrophysical objects classified on multiple,
and often independent sets of criteria in different energy bands, but are believed
to be powered by an accreting super-massive black hole (SMBH) residing in the core
\citep{2008NewAR..52..227T}. They are mainly characterized by high luminosity and
 persistent variability across the entire accessible electromagnetic spectrum,
especially at high energies ($\gtrsim$ X-rays). The observed trends/features across 
spectral, temporal and spatial domains are highly energy dependent and can manifest at
widely different scales in different energy bands. Despite the broad range of manifestations,
AGNs, under the simplest unification scheme based on observable characteristics,
have been primarily categorized on the  basis of radio-loudness and the properties
of optical-ultraviolet (UV) emission lines \citep{1995PASP..107..803U,2008NewAR..52..227T}.
The division is fully empirical with further hierarchical classification within
each class based on other observed features. Observationally, AGNs are the most
persistent but dynamic high energy emitters and thus, constitute a major fraction
of extragalactic objects in surveys. At \emph{Fermi}-LAT (Large Area Telescope)
$\gamma$-ray energies, they constitute $\sim$70\% of the total sources in the 3rd
\emph{Fermi}-LAT catalog \citep{2015ApJS..218...23A}.

The radio-loud AGNs, in general, are more dynamic than their corresponding radio-quiet
(RQ) counterparts and exhibit extreme scales in all spectral, temporal, spatial,
and polarization domains. Most prominently, they exhibit large scale (kpc/Mpc),
highly collimated relativistic jets which are the primary features for further
classifications  based on its orientation and luminosity. Broadly, if the jet
is orientated at close angles to the observer's line of sight, they are classified
as blazars,  and are called radio galaxies otherwise. Further, within radio-galaxies,
the low and high power sources are known as the Fanaroff and Riley I (FR I) and
FR II objects \citep{1974MNRAS.167P..31F}, respectively. The corresponding
counterparts in the blazar subclass  are, respectively, the BL Lacertae objects (BLLs)
and the flat spectrum radio quasars (FSRQs), where the former show  weak emission
lines or a featureless optical-ultraviolet (UV) spectrum while the latter exhibits
prominent broad emission lines \citep{1995PASP..107..803U}. 

One of the important characteristics of accretion-powered sources, regardless of their
mass and  compactness, is the high, frequent and energy-dependent temporal variability
e.g. AGNs \citep[e.g.][]{2004ApJ...612L..21G, 2004MNRAS.348..783M, 2005MNRAS.359.1469M},
galactic X-ray binaries \citep[GXBs; e.g.][]{2009ApJ...697L.167G}, micro-quasars
\citep[Cyg X-1;][]{2005MNRAS.359..345U}, cataclysmic variables and young stellar
objects \citep[YSOs;][]{2015SciA....1E0686S, 2015MNRAS.448.2430V,2012MNRAS.421.2854S}.
Such sources are generally characterized by their power spectral density (PSD),
flux distribution and excess-variability (RMS), and often show a (broken)
power-law PSD, a log-normal flux distribution, and a linear rms-flux relation.
These features, claimed as universal characteristics of accretion-powered
sources \citep[e.g.][]{2015SciA....1E0686S, 2012MNRAS.421.2854S},
 have also been observed in the emission of  the most under-luminous SMBH Sgr A$^\ast$
located at the center of our galaxy \citep{2014ApJ...791...24M, 2012ApJS..203...18W,
2011ApJ...728...37D, 2009ApJ...694L..87M}. These phenomenological similarities across
a wide range of mass scales ($\sim$ 10 orders) have further led
to the claim that the physics of accretion is universal, irrespective of accretor
physical details. 

At high energies in radio-loud sources, the primary contribution is believed to be
mainly from the inner accretion disk, corona, and the relativistic jets. 
Since in RQAGNs and GXBs, the accretion disk (inner region) primarily emits in X-ray
regime, a leading explanation \citep{2005MNRAS.359..345U, 2005MNRAS.359.1469M} for
the scale-free but similar behavior has been fluctuations in the accretion disk
\citep{1997MNRAS.292..679L}. The recent finding of similar behaviors in optical
observation of compact objects like CVs and non-compact objects like YSOs seems
consistent with this understanding as the accretion disk emission for these sources
lies in optical/UV. Additionally, similar characteristics have been seen in the
optical emission of few GXBs \citep{2009ApJ...697L.167G}. However, on a completely
different scale, the X-ray emission from the  whole Solar surface shows similar behavior
\citep{2007HiA....14...41Z} where the variability is well understood to be
a manifestation of magnetic reconnection process. Moreover, all accreting compact
systems are believed to have magnetic fields of varying degree, making the 
uniqueness claim ambiguous  \citep[see, for instance, a possible magnetic
reconnection scenario for flares from microquasars and non-blazar low luminous AGNs
in][]{2005A&A...441..845D,2015ApJ...802..113K,2015MNRAS.449...34K,2015ApJ...799L..20S}. In blazars,
almost all the emission, on the other
hand, is understood to be originating primarily in the jet as a result of the interaction
of jet plasma and magnetic field. Though a strong, large-scale magnetic field
is believed to be the primary driver of jet formation and collimation, it is not
clear whether only the BH spin, the accretion disk or both play a dynamic role in
its origin and evolution. Further, current studies suggest that even small scale magnetic
fields can efficiently power the jets \citep{2015MNRAS.446L..61P,2015ApJ...805..105C}.

Time series analysis and characterization is a powerful tool to understand highly
variable and unresolved sources and has been routinely used to explore the X-ray
emission from AGNs and GXBs along with spectral studies. Similar attempts to
statistically characterize $\gamma$-ray emission from blazars have been hampered due to
the lack of continuous detection and inhomogeneous temporal  coverage. This has been
a major impediment, making the results and interpretation prone to biasing
\citep[e.g.][]{2010A&A...520A..83H,2009A&A...503..797G,2016ApJ...822L..13K}.

In this work, we present the first systematic characterization of long-term $\gamma$-ray
emission from four AGNs:  BL Lac Mrk421 (z = 0.030), FSRQ B2 1520+10 (z = 1.487)
and PKS 1510-089 (z = 0.36), and FR I radio galaxy NGC 1275 (z = 0.017559) using
histogram, rms-flux relation, and PSD. An almost continuous detection ($> 97\%$)
of these sources over a data time-bin of 3-days makes them  ideal candidates to
study these features and comparison vis-a-vis other AGNs and  GXBs. The $\gamma$-ray
being dominated by the jet carries the imprints of jet processes and magnetic field
on the characteristics of the central source that,  otherwise, could have been observed
had the jet emission not overshadowed it. Thus, the similarity (dissimilarity) of
features should contain important physics about the relative role of the central
source features and magnetic field and/or jet processes, offering new insight into
the central source and jet connection. In the next
section, we present the details of the data and reduction procedures. In Section 3,
the temporal and flux variability  is presented along with the results. Implications
and connections with AGNs and X-ray binaries are discussed in Section 4 with
final  conclusions in Section 5.

\section{DATA REDUCTION}\label{sec:data}
The LAT is an electron-positron pair conversion imaging instrument on board the
\emph{Fermi Gamma-ray Space Telescope}.  It normally operates in scanning mode
and covers the entire sky for $\gamma$-ray photon events from 20 MeV to $>$ 300 GeV
every $\sim 3$ hours with a field of view of $\sim 2.4$ steradian \citep{2009ApJ...697.1071A}.
 It thus provides an almost uniform and continuous  observation of both, persistent
as well as transient $\gamma$-ray sources. The continuous survey since its launch
in 2008 has resulted in one of the longest and most evenly sampled $\gamma$-ray
data of the sky. Here, we have used the latest, \emph{PASS8} processed LAT data of
four AGNs from August 05, 2008 (MJD 54683) to October 05, 2015 (MJD 57300) to generate
the light curves over a time bin of 3-days. The choice of 3-day time bin is
to minimize the non-detections at shorter time-bins which can bias the histograms
and at the same time, provide maximum data cadence for the rms-flux and PSD analysis.

The analysis was performed with the latest \emph{Fermi Science Tool (v10r0p5)} following
the recommended procedure documented online\footnote{http://fermi.gsfc.nasa.gov/ssc/data/analysis/scitools/python\_tutorial.html}.
For each time-bin, events tagged as ``evclass=128, evtype=3'' with energy $> 100$ MeV
were selected from a circular region of interest (ROI) of $\rm 15^\circ$ centered
on the source. The events associated with the Earth's limb were minimized by
restricting the acceptance within a maximum zenith angle of $\rm 90^\circ$ while
the instrument operation in scientific data mode was ascertained by the recommended
time interval filter expression ``(DATA\_QUAL$>$0)\&\&(LAT\_CONFIG==1)''. The effect
of the selections and cuts along with the contributions from other $\gamma$-ray
sources in the selected field were incorporated using an exposure map over ROI and
an additional annulus of $\rm 10^\circ$ around it. The contribution of Galactic diffuse
and isotropic extragalactic emission was accounted by the inclusion of respective
emission templates \emph{gll\_iem\_v06.fits} and \emph{iso\_P8R2\_SOURCE\_V6\_v06.txt}
provided by the LAT Science Team in the input model file. The resulting files were
then modeled using an unbinned likelihood analysis method provided with the pylikelihood
library of the analysis software.

The above procedure is followed with $\gamma$-ray emission from other sources in
the field accounted for by an input model file from the LAT 3rd catalog \citep[3FGL
-- gll\_psc\_v16.fit;][]{2015ApJS..218...23A} describing their spectrum. In the
analysis, Mrk 421 is modeled as a power-law spectrum while a log-parabola spectral model
 has been used for the other three, with normalization and spectral
indices free to vary during the likelihood analysis.  Besides these, other sources
have been modeled as per their default spectra in the  3FGL catalog. The likelihood
convergence was performed iteratively by removing point sources having test statistic
(TS) $<$ 0 as done in \citet{2014ApJ...796...61K}. However, for detection and scientific
analysis, only data points which satisfy the significance of $\gtrsim$ 3$\sigma$,
corresponding to a TS value of $\gtrsim$ 9 are considered.

\section{ANALYSIS AND RESULTS}\label{sec:result}
\begin{figure}
 \centering
\includegraphics[scale=1.,angle=0]{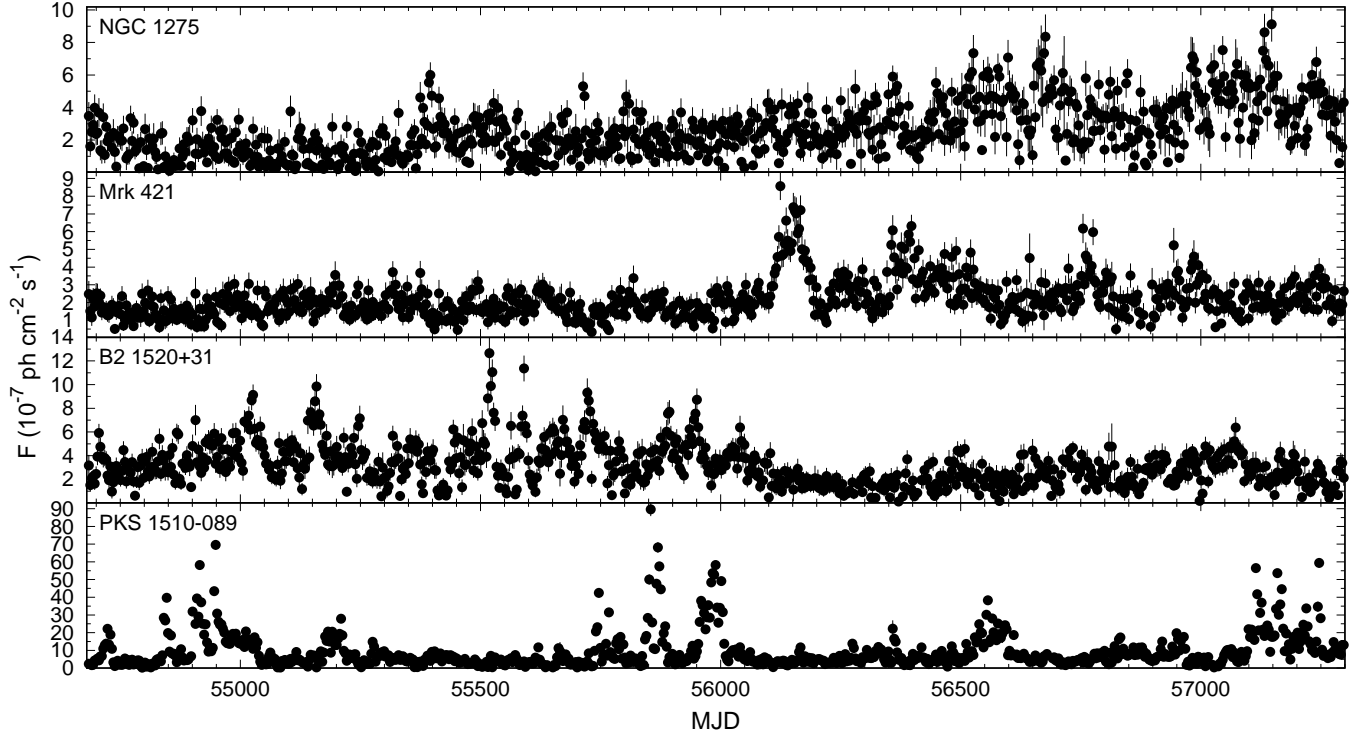}
\caption{The 0.1-300 GeV LAT $\gamma$-ray light curves of AGNs: NGC 1275, Mrk 421,
B2 1520+31 and PKS 1510-089 over an data integration time of 3 days from August 05,
2008 to October 05, 2015.}
\label{fig:latLc}
\end{figure}

The 0.1-300 GeV $\gamma-$ray light curves of the four AGNs: the FRI radio galaxy NGC 1275,
BLL Mrk 421, FSRQ B2 1520+31, and FSRQ PKS 1510-089 between August 05, 2008 (MJD 54683)
to October 05, 2015 (MJD 57300) over a uniform integration time of 3-days with
detection significance above/equal $3\sigma$ are shown in Figure \ref{fig:latLc}.
All the sources are detected over $97 \%$ \citep[e.g. PKS 1510-089;][]{2016ApJ...822L..13K}
of the total duration, thereby presenting one of the best sampled $\gamma$-ray light
curves till date (on the timescale of days). The light curves clearly show different levels
of variability for each source, an aspect of source and energy dependent variability.
We explore the statistical features and quantify this in terms of flux and temporal variability
using flux-histogram, RMS and PSD\footnote{The scope of these statistical methods
to explore and understand AGNs and other accreting compact sources from variability
point of view is outlined e.g. in \citet{2003MNRAS.345.1271V,2005MNRAS.359.1469M,
2005MNRAS.359..345U,2001MNRAS.323L..26U}}.

\subsection{Flux Distribution and Effect of Photon Statistics} \label{subsec:hist}
\begin{figure}
\centering
\includegraphics[scale=1.,angle=0]{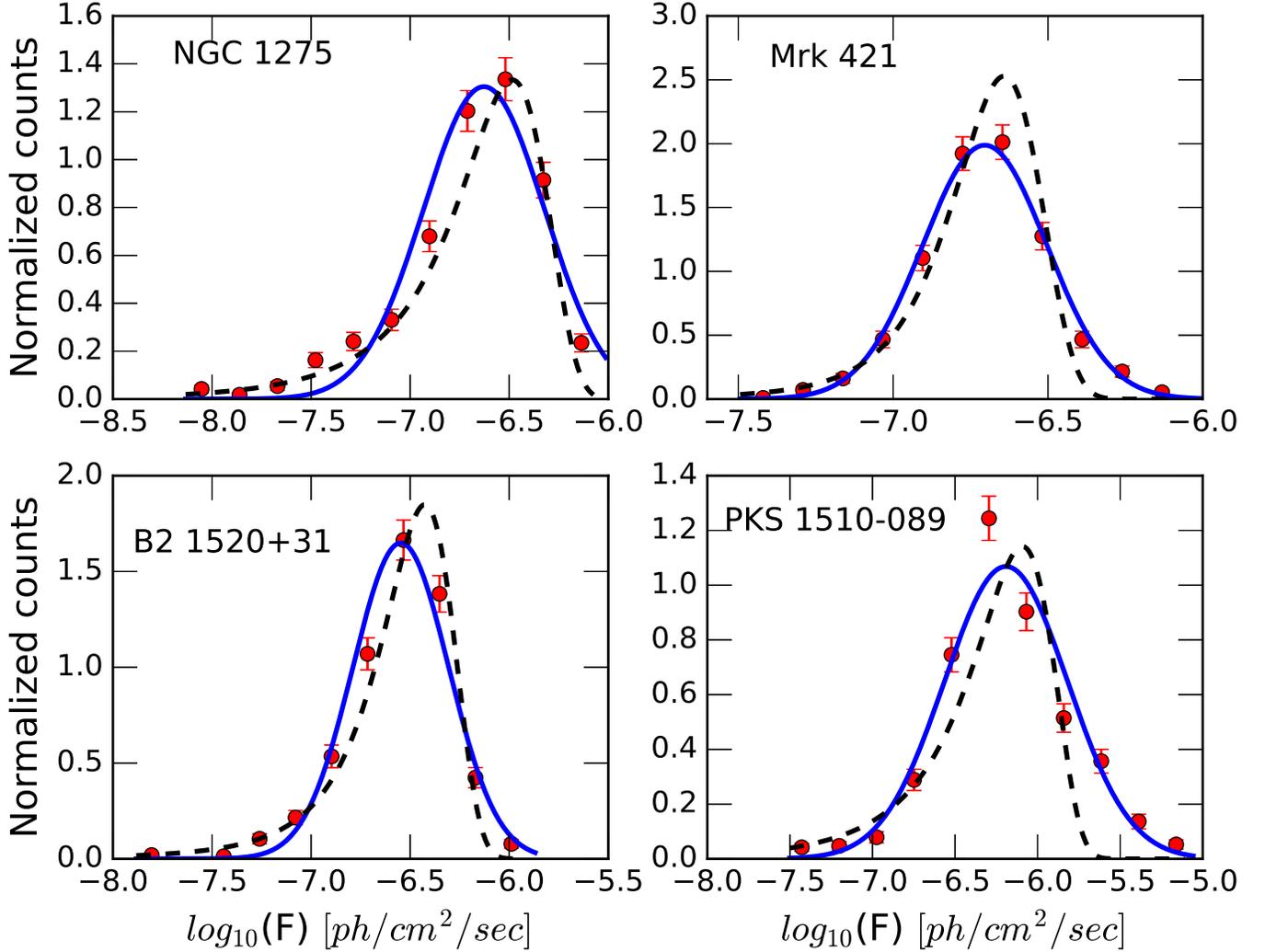}
\caption {Normalized histograms of logarithmic $\gamma$-ray photon flux fitted with
a lognormal (solid blue line) and Gaussian (dotted black line) (see \S\ref{subsec:hist}).}
\label{fig:logFlux}
\end{figure}

A histogram of a quantity characterizes the distribution of the measured values
and  thus is an important tool to explore the existence of characteristic scales/values in a
system and the range it spans. In Figure \ref{fig:logFlux}, we present the
normalized histograms of the logarithmic\footnote{Unless stated otherwise, logarithm
throughout this work means log of physical quantity to the base of 10.} $\gamma$-ray
photon flux. Care was taken to ensure a bin width larger than the  flux value
with account of error in that bin. All the histograms show a prominent peak and a low flux tail with flux
spanning two  orders of magnitudes between the extremes except for Mrk 421. Further,
the peaks have a typical variation of about an order of magnitude which is quite
common in PKS 1510-089 followed by NGC 1275, B2 1520+31 and Mrk 421. Since a
random stochastic process is expected to yield a normal distribution, we computed the skewness
of the distributions (Table \ref{tab:histFit}). While the skewness deviates from
zero for the three blazars, the deviation for the  radio-galaxy is substantially less. We then
thus checked the consistency of the distributions with a lognormal as is normally
inferred for RQAGNs, GXBs and even blazars \citep{2016ApJ...822L..13K,2010A&A...520A..83H,
2009A&A...503..797G}. The skewness of the logarithm of the fluxes was  also computed, which
was found to be  closer to zero for the three blazars, but significantly large
for the  radio-galaxy.  Subsequently, we computed the Anderson Darling test
statistics \citep[nortest\footnote{\url{https://cran.r-project.org/package=nortest}};][]{R2013}
for each of the light-curves, as coming from a Gaussian or a lognormal
distribution (Table \ref{tab:histFit}). While both distributions are strongly
disfavored for NGC 1275, a lognormal is preferred for the three blazars. To
further understand the distributions, we fitted them with normal and lognormal
functions and compared the results using a $\chi^2$ test.
The results of the best fit values are given in Table \ref{tab:histFit} and the plot
is shown in Figure \ref{fig:logFlux}. The results are quite interesting favoring the
lognormal distributions for the 3 blazars,  Mrk 421, PKS 1510-089 and B2 1520+31,
while the distribution for NGC 1275 seems  more complex where neither matches the observed data.

At $\gamma$-ray energies with lesser and lesser photons at higher energies, photon
statistics play a vital  role. A significant error due to photons statistics
can bias the intrinsic variation as well as the low flux end in the histogram. To understand the
effects of errors on the flux and observed variations, we generated two intrinsic
distributions: a lognormal and a Normal flux distribution  and then convolved
each with Poisson noise to reproduce the observed mean and widths. Again, for all
the blazars, an intrinsic log-normal well reproduces the observed data.  However,
while for NGC 1275 an intrinsic Gaussian is favored over an intrinsic log-normal,
the underlying distribution seems more complex than either of the distributions
studied in the present work\footnote{a detail study using a sample of blazars is
under preparation (Z. Shah et al)}.

\begin{table}
\centering
\caption{Fit statistics for the distributions}
\begin{tabular}{l c c c c c c c}
\hline \hline 
Source	& \multicolumn{6}{c}{log(Flux)} \\ \cline{2-8}
Name 	& mean & $\sigma$ & $\chi_{red}^2$ & dof & $\rm \kappa$ & $\rm \langle{df/f/ln(10)}\rangle$ & AD (prob)\\
\hline
NGC 1275 & -6.63 $\pm$ 0.03 & 0.30 $\pm$ 0.03 & 7.9 & 9 &  -1.13 & 0.15  & 11.8 ($<2\times10^{-16}$) \\
Mrk 421	& -6.70 $\pm$ 0.01 & 0.20 $\pm$ 0.005 & 1.2 & 9 & -0.13 & 0.12 & 1.1 ($4\times10^{-3}$) \\
B2 1520+31 & -6.54 $\pm$ 0.02 & 0.24 $\pm$ 0.01 & 2.6 & 8 & -0.61 & 0.12 & 5.4 ($1\times10^{-13}$)  \\
PKS 1510-089 & -6.19 $\pm$ 0.02 & 0.37 $\pm$ 0.02 & 2.6 & 9 & -0.08 & 0.08 & 3.7 ($2\times10^{-9}$) \\
\hline\hline
	     &  \multicolumn{6}{c}{Flux} \\ \cline{2-8}
	     & mean$^*$& $\sigma^*$ & $\chi_{red}^2$ & dof & $\rm \kappa$ & $\rm \langle{df^*}\rangle $ & AD (prob) \\
\hline
NGC 1275     & 2.16 $\pm$ 0.21 & 1.90 $\pm$ 0.18 & 4.7 & 9 & 0.76 &0.72 &9.4 ($<2\times10^{-16}$)\\
Mrk 421      & 2.01 $\pm$ 0.09 & 7.8 $\pm$ 0.07 & 11.44 & 9 & 1.67 &0.71 & 20.5 ($<2\times10^{-16}$) \\
B2 1520+31   & 2.99 $\pm$ 0.18 & 1.68 $\pm$ 0.15 & 7.73 & 8 &1.16 &0.52 & 9.3 ($<2\times10^{-16}$) \\
PKS 1510-089 & 4.65 $\pm$ 0.12 & 5.25 $\pm$ 0.09  & 18.5 & 9 & 2.95 &1.18 & 76.3 ($<2\times10^{-16}$) \\
\hline
\hline
\multicolumn{7}{l}{f: flux (ph cm$^{-2}$ s$^{-1}$); $^*$: $\times 10^{-7}$ (ph cm$^{-2}$ s$^{-1}$)}\\
\multicolumn{7}{l}{$\rm \kappa$: skewness {(\bf $1\sigma$ error $\pm0.13$)}; $\rm \chi^2_{red} = \chi^2/dof$; dof: degree of freedom}\\
\multicolumn{7}{l}{$\rm AD$: Anderson Darling test statistics}\\
\end{tabular}
\label{tab:histFit}
\end{table}

\subsection{Gamma-ray PSD}\label{subsec:psd}

\begin{table}[!htbp]
\centering
\caption{Best Fit parameters for the PSDs}
\begin{tabular}{l c c c c c c c c c c}
\hline \hline 
Source	& \multicolumn{3}{c}{Power-law$^{\ref{fn:pl}}$} && \multicolumn{5}{c}{Broken Power-law$^{\ref{fn:bpl}}$} \\ \cline{2-4} \cline{6-10}
Name 	& 	$\Gamma$	& $\chi_{red}^2$ & dof && $\rm \Gamma_1$ & $\rm \Gamma_2$ & $\rm log(\nu_0)$ & $\rm \chi_{red}^2$ & dof \\ 
\hline
NGC 1275 & -0.96 $\pm$ 0.10 & 2.06 & 8 	&& -0.78 $\pm$ 0.10 & -1.98 $\pm$ 0.77 & -7.23 $\pm$ 0.35 & 1.09 & 6 \\
Mrk 421 & -1.20 $\pm$ 0.11 & 3.36 & 8 	&& -0.57 $\pm$ 0.48 & -1.48 $\pm$ 0.20 & -6.30 $\pm$ 0.31 & 2.39 & 6 \\
B2 1520+31 & -1.15 $\pm$ 0.09 & 1.80 & 8	&& -1.34 $\pm$ 0.23 & -0.87 $\pm$ 0.34 & -6.74 $\pm$ 0.65 & 1.89 & 6 \\
PKS 1510-089 & -1.06 $\pm$ 0.09 & 1.10 & 8 && -1.32 $\pm$ 0.12 & -0.39 $\pm$ 0.31 & -7.09 $\pm$ 0.22 & 0.54 & 6 \\
\hline
\hline 
\end{tabular}
\label{tab:fitParPSD}
\end{table}

\begin{figure}
\centering
 \includegraphics[scale=1.3]{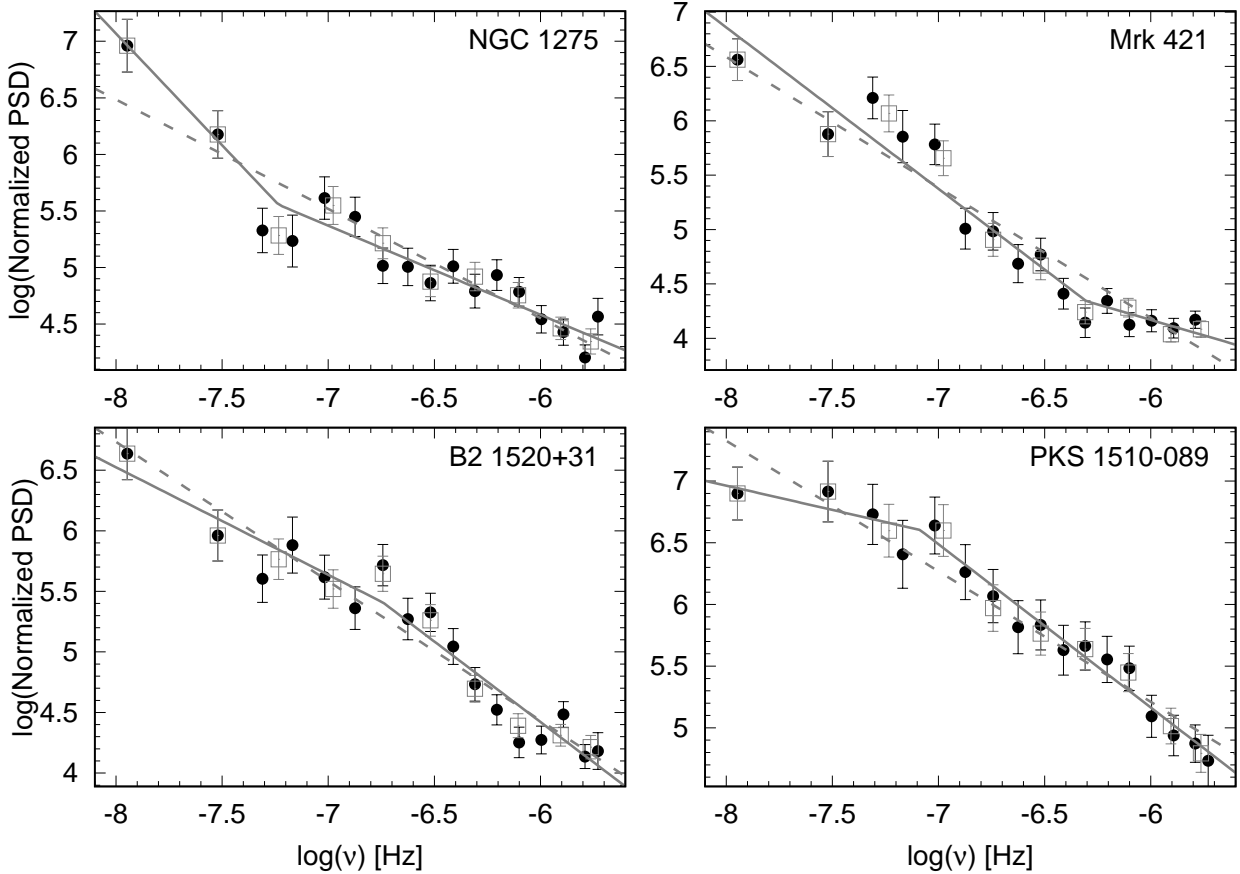}
\caption{Normalized PSDs with two different binning criterion (see \S\ref{subsec:psd}).
The grey points correspond to a binning criteria that is twice as that of the one
used for the black points. The dashed and solid gray curves are the best-fit power-law
and broken power-law to the gray  data, respectively (see Table \ref{tab:fitParPSD}).}
\label{fig:latPSD}
\end{figure}

PSD  characterizes the variability power of a source  and also  searches for the
presence of any characteristic timescale within the duration of interest. In
Figure \ref{fig:latPSD}, we present the binned normalized PSDs following the method outlined
in \citet{2005astro.ph..1215G}, for all the four AGNs with two different bin significances.
The binning significance of the gray points are twice as  that used for the black
data points. The error-bars on the PSDs were estimated by simulating light curves
for each source  with its observed PSD using the method of \citet{1995A&A...300..707T}.
For this, we simulated 1000 realization of light curves for each source and sampled each
according to the observed light curve. We then estimated the PSD for each  one as mentioned
above and binned it using bins of the PSD of the observed light curve. The width of the
distribution of the simulated PSDs then provides the error bar for the observed PSD.
 We fitted these with both a power-law\footnote{$\rm f(\nu) \propto \nu^\Gamma$ \label{fn:pl}}
and a single broken power-law model\footnote{$\rm f(\nu) \propto \nu^{\Gamma_1}~ (\nu < \nu_0)
~\& \propto \nu^{\Gamma_2}~ (\nu \geq \nu_0)$ \label{fn:bpl}} and the respective
best-fit values are reported in Table \ref{tab:fitParPSD}. It should be noted that
the broken power-law  suggests different break timescales. However, we have quoted
only the break which resulted in the least reduced-$\chi^2~(\rm \chi^2_{red})$ value.

\subsection{RMS Variability}\label{subsec:rms}
\begin{figure}
\centering
\includegraphics[scale=1.4,angle=0]{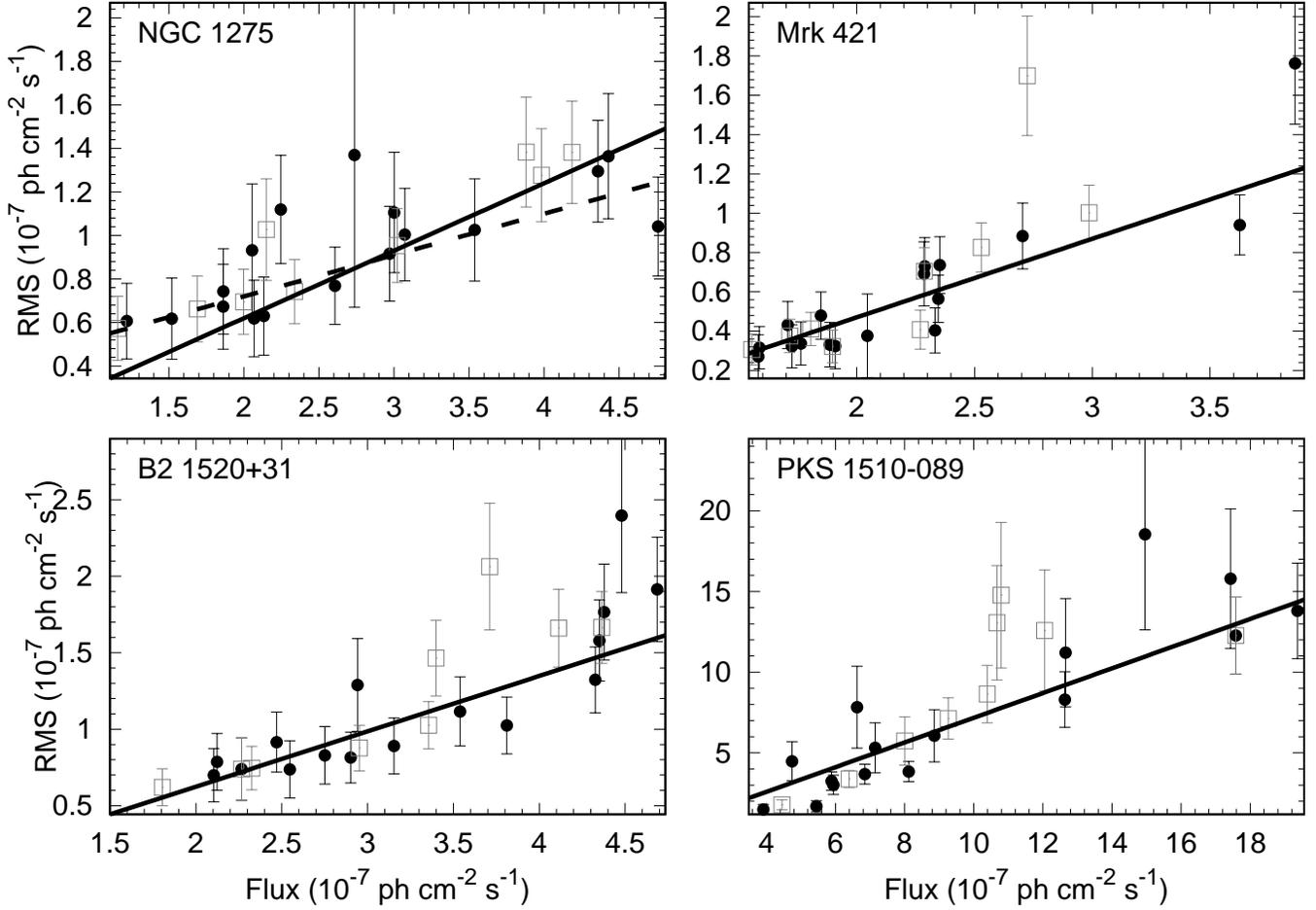}
\caption{RMS-flux variability (not normalized) of the four LAT AGNs. The black and grey
points correspond to 50 and 100 data-points per bin (see \S\ref{subsec:rms}) while the
black solid line is the best fit linear relation to the black data points (see Table
\ref{tab:fitParRMS}) while the dotted curve for NGC 1275 is fit with intercept
fixed to zero.}
\label{fig:latRMS}
\end{figure}

\begin{figure}
\centering
\begin{minipage}[t]{0.485\linewidth}
\includegraphics[scale=0.7,angle=0]{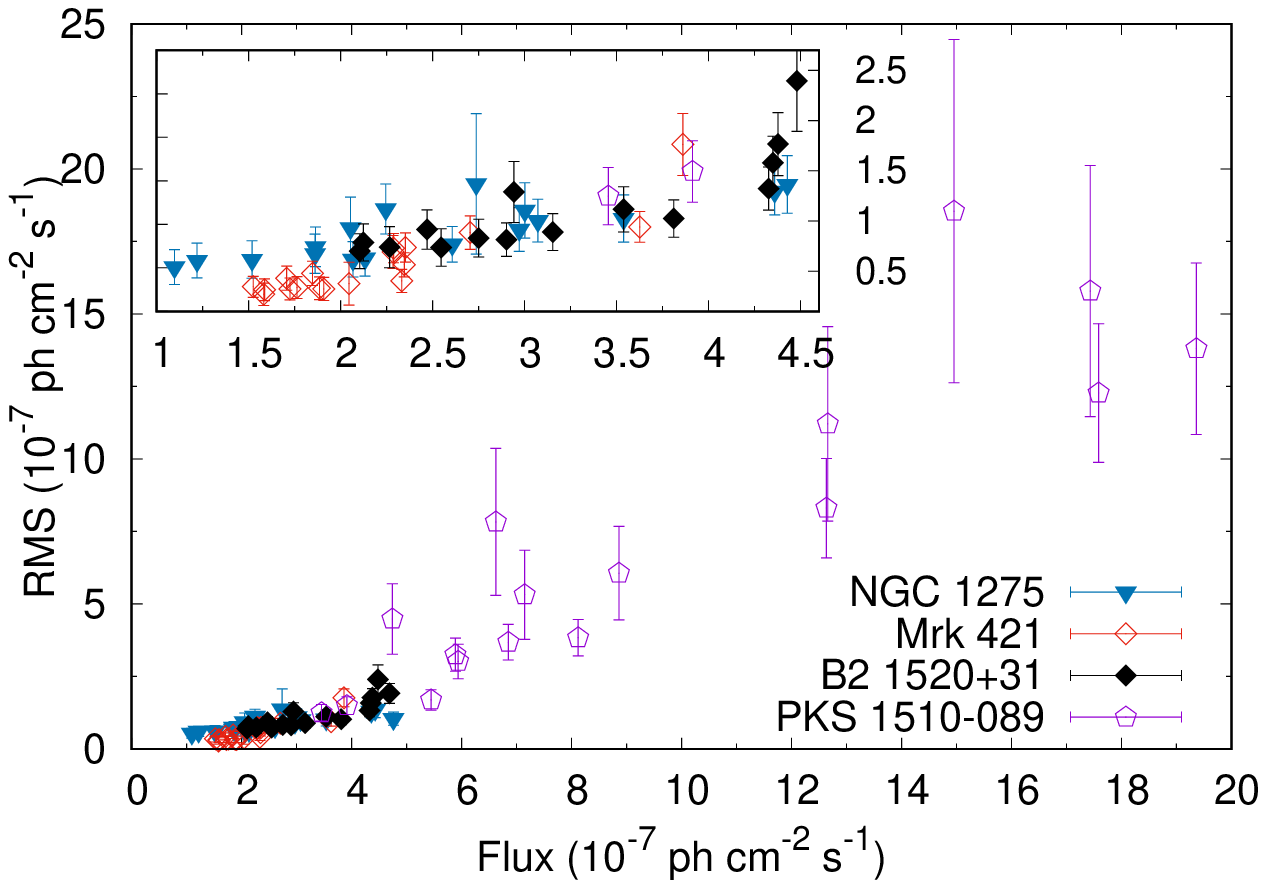}
\end{minipage}
\centering
\begin{minipage}[t]{0.485\linewidth}
\includegraphics[scale=0.7,angle=0]{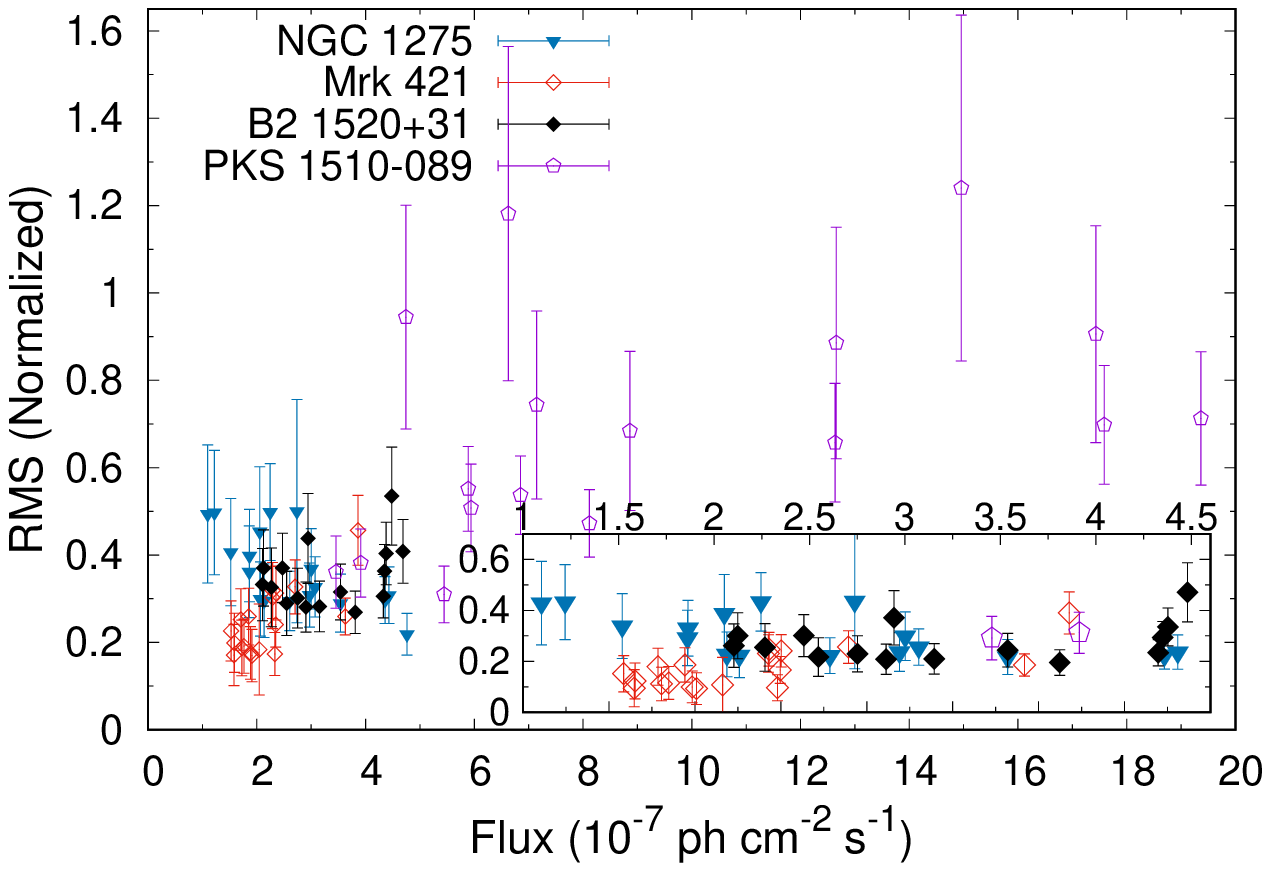}
\end{minipage}
\caption {RMS-flux relation with 50 data per bins for all the sources on the same
scale in left and the normalized one on the right with zoomed plot in the inset
at low flux  values.}
\label{fig:minipageRMS}
\end{figure}

To quantify and compare the degree of intrinsic variability, we have estimated the
rms-flux relation for two different binnings of 50 and 100 data per bin. The
binning is according to the time-series obtained i.e. time. The estimates are
shown, respectively, in black and gray data points in Figure \ref{fig:latRMS}. The
error-bars consist of both the effect of observed errors and true variability of
the source added in quadrature. The later is calculated by simulating 1000
realizations of light curves with 50 and 100 data points for each source and then
estimating the rms \citep{2003MNRAS.345.1271V}. We also show the best fit linear
model to the rms-flux relation with the fit values reported in Table \ref{tab:fitParRMS},
along with the Spearman correlation and the null hypothesis probability. In Figure
\ref{fig:minipageRMS}, we show the rms-flux relation from 50 data per bins for
all the sources on one scale in the left panel and the same but with normalized-rms
on the right.

\begin{table}[htb!]
\centering
\caption{Best Fit parameters for a linear Flux-RMS relation}
\begin{tabular}{l c c c c c c c}
\hline \hline 
Source   & \multicolumn{4}{c}{RMS = m*Flux+c} && \multicolumn{2}{c}{$\rm Spearman~ \rho~ (probability)$}\\ \cline{2-5} \cline{7-8}
	&  	m & c & $\rm\chi_{red}^2$ & dof	&& 50 data/bin & 100 data/bin \\ \hline
\multirow{2}{*}{NGC 1275} &  0.19 $\pm$ 0.03 & (+3.4 $\pm$ 0.8)$\times 10^{-8}$ & 0.38 & 16 && 0.81 ($4.6\times10^{-5}$) & 0.90 ($9\times10^{-4}$) \\
	&  0.32 $\pm$ 0.02 & 0.0  & 0.78 & 17 && 0.81 ($4.6\times10^{-5}$) & 0.90 ($9\times10^{-4}$) \\
Mrk 421  &  0.40 $\pm$ 0.06 & (-3.3 $\pm$ 1.2)$\times 10^{-8}$ & 0.85 & 16 && 0.83 ($1.6\times10^{-5}$) & 0.90 ($9\times10^{-4}$) \\
B2 1520+31 &  0.36 $\pm$ 0.06 & (-1.1 $\pm$ 1.8)$\times 10^{-8}$ & 0.78 & 14 && 0.60 ($1\times10^{-2}$) & 0.95 ($8.8\times10^{-5}$) \\
PKS 1510-089 &  0.76 $\pm$ 0.09 & (-1.6 $\pm$ 0.5)$\times 10^{-7}$ & 1.40 & 15 &&  0.90 ($6.3\times10^{-7}$) & 0.85 ($3.7\times10^{-3}$) \\
\hline
\hline 
\end{tabular}
\label{tab:fitParRMS}
\end{table}

\section{DISCUSSION}
The range of AGNs behavior in spectral, temporal and spatial domains is huge and
diverse, reaching extremes in blazars and sources exhibiting blazar-like characteristics.
Studies in time domain over different timescales at $\gamma$-ray energies show that
the variability is stochastic \citep{2013ApJ...773..177N, 2014ApJ...786..143S} with
variations on all scales from minutes, days to years. Being dominated by the jet
emission, imprints from central engine evolve in a non-linear fashion along the
jet, leading to varying temporal features and giving rise to an energy dependent
variability. Even the most prominent characteristic feature, the broad double-humped
shaped SED shows substantial evolution on different timescales and has been
used extensively to explore  different scenarios of emission processes. On the other
hand, a broader understanding of physical processes can be obtained by a 
careful and systematic investigation of  the statistical properties in time domain of
well selected sources over long  timescales \citep[e.g.][]{2016ApJ...822L..13K}
by comparing with current blazars physics as well as with  non-blazar counterparts
and with other accretion-powered sources.

In a first such attempt at $\gamma$-ray energies, we have characterized the emission
from four AGNs: FR I radio-galaxy NGC 1275, BLL Mrk 421, FSRQ B2 1520+31, and
FSRQ PKS 1510-089, using one of the best-sampled light curves to date on timescale
of days  through their flux histograms, rms-fluxes and PSDs, which  are used to
explore and understand other accretion powered sources like RQAGNs, GXBs, CVs etc
at X-rays/UV/optical energies. An important caveat in this analysis is the assumption
of stationarity in the light-curves. However, since the PSD power-spectral
index is less than $2$ in all the cases, this assumption is valid. Moreover, we
computed the Augmented Dickey-Fuller test statistics for the null hypothesis that
these are non-stationary time series, and the null hypothesis could be rejected
with a significance level of less than 1\% for each of the light curves.

The systematic investigation performed here reveals a broad range of statistical features with
sources exhibiting a flux variability spanning $\gtrsim 10^2$ between the extremes
(barring non-detections which are $\lesssim 3\%$) except for Mrk 421 for which it
is $\sim 10$. All the histograms exhibit a prominent peak with a typical width of
an order of magnitude in terms of flux  variability, and an extended low flux tail.
For the three blazars, a lognormal fit describes the histogram better
compared to a Gaussian, while NGC 1275 distribution is more complex than either
(see Table \ref{tab:histFit}). Interestingly, albeit possessing very different
duty cycles, all the sources are almost equally variable ($\sigma$, Table \ref{tab:histFit})
in the LAT energy band. A different variability of PKS 1510-089 is due the
fact that its flux distribution is bi-lognormal with equal variability in both
\citep{2016ApJ...822L..13K} and consistent with values inferred for others here. 

Though the histograms are skewed/lognormal, the rms-flux relation is linear for all
the sources with large scatters and a flux offset different from zero at the zero rms.
Excepf for FR I source NGC 1275, the offset (= -c/m, Table \ref{tab:fitParRMS}) is 
positive for the others, i.e. the blazars. The negative flux
offset for NGC 1275 may be due to random Poisson fluctuations at low fluxes
\citep[e.g.][]{2007HiA....14...41Z}. Fixing the offset to zero for NGC 1275
leads to a slope value consistent with the two blazars, Mrk 421 and
B2 1520+31. However, this remains very different from PKS 1510-089 which is more
variable at a given flux level compared to the others (Fig \ref{fig:latRMS} and Table
\ref{tab:fitParPSD}). The PSDs, on the other hand, are typical of accretion-powered
compact sources showing a scale-free power law profiles over the duration of the
study with hints of break (for B2 1520+31 and PKS 1510-089). Such breaks have
been clearly observed in the X-ray emission of many AGNs and GXBs and have
been found to be correlated with the BH \citep{2005MNRAS.359.1469M,2006Natur.444..730M}.
Here, however, the quality of the data and large scatter suggests many
break points and/or break at different frequencies in the case of a single
broken power-law fit. Further, the PSD profile in the case of NGC 1275 and Mrk 421
shows rising at long timescales, suggesting increased variability and thus,
more power,  contrary to expectation, while in the other two, the variability  shows
decrement but the broken power-law does not improve the fit significantly. Thus,
within the duration considered here, the PSDs suggest no scale/break \citep[Table
\ref{tab:fitParRMS}, see also][]{2013ApJ...773..177N}.

Many models of blazars variability  have been proposed in the literature, e.g.
shock-in-a-jet \citep{1985ApJ...298..114M,2010ApJ...711..445B}, turbulent extreme
multi-zone \citep{2014ApJ...780...87M}, needle-in-a-jet \citep{2008MNRAS.386L..28G},
jets-in-a-jet model \citep{2009MNRAS.395L..29G}. The latter, i.e.  jets-in-a-jet
model is also proposed for the $\gamma$-ray emission from radio galaixes \citep[e.g.
M 87;][]{2010MNRAS.402.1649G}. Though these models explain the spectral and
temporal features at different epochs, a statistical study of these features
has not been performed except for the minijets-in-a-jet model \citep[see also \citet{2012MNRAS.426.1374C}]{2012A&A...548A.123B}.
These much needed theoretical investigations involving multiple parameters are
beyond the scope of present work and are an extensive study in themselves. Thus, we
discuss the results in the context of minijets-in-a-jet model and studies where these 
methods have been used to explore the physics of different sources.

In the minijets-in-a-jet model, emission from identical, independent,
but randomly oriented minijets follows a Pareto\footnote{
characterized by a probability density function, f(x) = a/x$^{1+a}$ \citep[e.g.][]
{2005PApGe.162.1187Z}} distribution which can produce a wide range of histograms
from a power-law  (in case of single region) to one consistent with a lognormal
(in the case of contributions from a large number of emission regions; e.g. see Fig.
4 \& 5 of \citet{2012A&A...548A.123B}), while preserving a linear rms-flux relation
in all the cases. This is what we have found in this study with histograms
having skewed/lognormal profiles following a linear rms-flux
relation and a constant positive flux offset for  all except NGC 1275. The
model thus proposed for observations of  rapid variability in blazars and radio
galaxies (e.g. M87) exhibiting blazar-like characteristics  is consistent with
their long term statistical properties as well.  Importantly, emission in this model
is result of additive contributions  of the underlying physical process, rather than
multiplicative as argued by \citet{2005MNRAS.359..345U} for X-ray emission in GXBs and
non-jetted AGNs.

With the current set of data, it is not possible to verify whether the high flux
end is a power-law.  Though the minijets-in-a-jet model appears to explain the long term
statistical characteristics, the short term stastical predictions seems contrary
to the observations during flares or high flux states. Being additive with identical
properties of minijets except for Doppler  boosting, two similar $\gamma$-ray flares
from the same source should have similar features in other energy bands, contrary
to the observations \citep[e.g.][]{2014ApJ...796...61K}. Further, often during
flares, the correlated multi-wavelength variability and light curve profiles
suggest emission dominated by a single emission region. However, most of the flaring
histograms are  consistent with a lognormal distribution  \citep[e.g.][]{2017MNRAS.464.2046K,
2010A&A...520A..83H}, contrary to the  statistical results  which resemble
lognormal only in the case of contribution from a large number of emission regions
\citep{2012A&A...548A.123B}. Additionally, the origin of different proportionality
constant for the rms-flux relation as found here needs further investigation and so
the PSDs from this model.

The lack of correlation between the mass of the central SMBH and the variability
time scale in blazars \citep[e.g.][]{2016ApJ...822L..13K}, contrary to what is inferred
in other non-jetted AGNs and GXBs \citep{2006MNRAS.372.1366K,2006Natur.444..730M,2005MNRAS.359.1469M}
already  suggests that the jet  can modify substantially the imprint of feedback from the
central engine, if any. However, the occurrences of extended periods of inactivity
(few months) at LAT $\gamma$-rays after a long activity period in powerful and highly
variable blazars (e.g. 3C 454.3) suggest that the trigger  probably depends on
feedbacks and ambient effects. The source of such feedback is most likely the energy
inputs from central engine. Further, the typical SED modeling assuming single/multiple
emission regions suggests  that the emission sites are kinetic energy dominated
\citep{2014Natur.515..376G}. But such modeling always fails to explain the
radio emission and attributes its origin to a much-extended region  compared to
the optical, X-rays, and $\gamma$-rays. On the other hand, a magnetic reconnection
model in  a relativistic fluid approach within the jet, based on inferences
from M87 observations suggesting a Poynting flux dominated jet \citep{2013ApJ...775...70H}
seems to better describe the blazars' SEDs from radio-to-GeV/TeV \citep{2015MNRAS.453.4070P}.
An alternative model suggests that turbulent magnetic reconnection in the coronal
region around the accretion disk  can explain the emission from compact accreting
sources spanning ten  orders in the BH mass, but fails for blazars and GRBs
\citep{2015ApJ...802..113K,2015ApJ...799L..20S} i.e.  sources with the jet pointing
to the line-of-sight, where jet overwhelms the nuclear emission.  Interestingly,
the statistics of physical quantities related to the Solar X-ray emission and
coronal mass ejections (CMEs), understood to be powered by magnetic-reconnection
\citep{2017ApJ...836...17A} also exhibit lognormal properties
\citep{2007HiA....14...41Z,2003ICRC....5.2729A} are consistent with our finding in this work.
Moreover, apart from emission and mass ejections, CMEs also lead to
energization and generation of non-thermal particles \citep[e.g.][]{2017ApJ...835..219A}. 
Though it is not clear how much of total energy is channeled into  the various
modes, i.e. emission, matter ejection, non-thermal and relativistic particles. 
Studies of jets via numerical simulations suggest that both small and large scale
fields are capable of powering jets \citep{2016ApJ...824...48S,2015MNRAS.446L..61P,2015ApJ...805..105C,
2014MNRAS.440.2185D,2009MNRAS.395.2183Y}. However, both jet physics and the 
nature of magnetic being steady or transient \citep[e.g.][]{2003Natur.426..533F}
are topic of current active investigations.

Lognormality in blazars have been inferred in many sources and in different energy
bands, e.g. X-rays \citep{2016ApJ...822L..13K,2009A&A...503..797G,2002PASJ...54L..69N}, $\gamma$-rays 
\citep{2016ApJ...822L..13K,2017MNRAS.464.2046K,ln1011,2016A&A...591A..83S,2010A&A...520A..83H},
optical and IR \citep{2016ApJ...822L..13K}. In RQAGNs, GXBs and other accretion-powered
sources, the lognormality implying a linear rms-flux relation is normally claimed
to be an imprint of multiplicative effect associated with the accretion disk \citep{2005MNRAS.359..345U,
1997MNRAS.292..679L}. Thus, the skewed/lognormal flux distribution with a linear flux
histogram inferred here and as predicted by the minijets-in-a-jet model
\citep{2012A&A...548A.123B}, in combination with highly magnetic systems and rapid
variability favors magnetic reconnection origin. Within this, the inferred statistical
results for NGC 1275 suggest contribution from a fewer emission regions compared
to the blazars. Further, given the jet is being powered by a compact source, the
reconnection may or may not reflect the imprint of the central engine. If the emission
is triggered as a result of fluctuations/perturbation from the central engine and
the jet channels the energy into non-thermal particles, then the observed features may
be an imprint of accretion-disk on the jet, as argued in \citet{2009A&A...503..797G},
where the efficiency of the non-linear processes in the jet gives rise to the wide
variations as normally observed. Second, the emission may be solely the result of
processes intrinsic to the jet without any direct correlation with the central engine
except for the energy input. The third potential possibility is if the $\gamma$-ray
emission from NGC 1275 is non-boosted from the disk corona, a likely location
of the jet launching region as argued in \citet{2015ApJ...802..113K} and \citet{2015ApJ...799L..20S},
then, similar imprints propagating downstream the jet may be boosted by the jet
bulk motion which, by virtue of Doppler beaming, also makes more regions to contribute
to the emission, making the flux distribution lognormal. It should be noted that a
Doppler boost ($\rm \delta$) of a few to ten can make the NGC 1275 flux ($\rm \propto
\delta^3/distance^2$) similar to the BL Lacs and FSRQs studied in this work, though
the energetics and dynamics details may be different for each source.

\section{CONCLUSION}
In this work, we explored the $\gamma$-ray variability characteristics of four
LAT AGNs: FR I NGC 1275, BLL Mrk 421, FSRQs B2 1520+31 and PKS 1510-089 with
all having a near continuous detection in \emph{Fermi}-LAT over an integration time
of 3-days. All the sources show huge variations in flux, ranging $\sim$ 1-2
order of magnitude between the extremes with PSDs consistent with a shot noise
powerlaw index of $\sim$ 1. For blazars, a lognormal profile describes flux
histogram better compared to a Gaussian while NGC 1275 flux distribution is complex
than either. Irrespective of the shape of the flux histogram, the rms-flux relation
is linear for all with a positive offset flux for blazars. These statistical features
are broadly consistent with the statistical predications of the reconnection powered
minijets-in-a-jet model as well as the characteristics of physical quantities
associated with the Solar X-ray emission and CMEs. The qualitative
similarity of features suggests that both the magnetic field and the jet matter
are tightly coupled, except at the highest flux states which  are generally believed
to be  emission from a single region compared to the collective contribution at
low flux levels. The results, also indicate that the magnetic reconnection
powered emission could be the result of the fluctuations from the accretion-disk,
or the jet dynamics, or even the imprint of disk corona, this later, 
particularly in the case of the non-blazar radio galaxy NGC 1275.

\section*{ACKNOWLEDGEMENT}
The authors thank the anonymous referee for invaluable comments and suggestions
that have improved the article.
PK thanks CB Singh for fruitful discussions. This work is partially supported
by grants from the Brazilian agency FAPESP 2015/13933-0. EMGDP also acknowledges
support from  FAPESP (2013/10559-5) and CNPq (306598/2009-4) grants. 
The \textit{Fermi}-LAT Collaboration acknowledges generous ongoing support from a number of agencies and 
institutes that have supported both the development and the operation of the LAT as well as scientific
data analysis. These include the National Aeronautics and Space Administration and the Department of 
Energy in the United States, the Commissariat $\grave{a}$ l'Energie Atomique and the Centre National de la Recherche
Scientifique/Institut National de Physique Nucl$\grave{e}$aire et de Physique des Particules in France, the
Agenzia Spaziale Italiana and the Istituto Nazionale di Fisica Nucleare in Italy, the Ministry of Education,
Culture, Sports, Science and Technology (MEXT), High Energy Accelerator Research Organization (KEK) and Japan
Aeros pace Exploration Agency (JAXA) in Japan, and the K. A. Wallenberg Foundation, the Swedish Research Council
and the Swedish National Space Board in Sweden. 
Additional support for science analysis during the operations phase from the following agencies is a
lso gratefully acknowledged: the Istituto Nazionale di Astrofisica in Italy and
and the Centre National d'Etudes Spatiales in France.

\facility{Fermi}
\software{Fermi Science Tool (v10r0p5), R \citep[\url{https://www.R-project.org}]{R2013},
Matplotlib \citep{Matplotlib}, gnuplot (\url{http://www.gnuplot.info})}


\end{document}